\documentclass[conference]{IEEEtran}
\IEEEoverridecommandlockouts
\usepackage{cite}
\usepackage{amsmath,amssymb,amsfonts}
\usepackage{algorithmic}
\usepackage{graphicx}
\usepackage{textcomp}
\usepackage{xcolor}
\def\BibTeX{{\rm B\kern-.05em{\sc i\kern-.025em b}\kern-.08em
    T\kern-.1667em\lower.7ex\hbox{E}\kern-.125emX}}
\begin{document}

\title{R v F (2025): Addressing the Defence of Hacking
}

\author{\IEEEauthorblockN{Junade Ali}
\IEEEauthorblockA{\textit{Engprax Ltd} \\
Edinburgh, Scotland, United Kingdom \\
junade@engprax.com}
}

\maketitle

\begin{abstract}
The defence of hacking (sometimes referred to as the ``Trojan Horse Defence'' or the ``SODDI Defence'', Some Other Dude Did It Defence) is prevalent in computer cases and a challenge for those working in the criminal justice system. Historical reviews of cases have demonstrated the defence operating to varying levels of success. However, there remains an absence in academic literature of case studies of how digital forensics investigators can address this defence, to assist courts in acquitting the innocent and convicting the guilty. This case study follows the case of \textit{R v F} where a defendant asserted this defence and the author worked alongside a police investigator to investigate the merits of the defence and bring empirical evidence before the jury. As the first case study of its kind, it presents practical lessons and techniques for digital forensic investigators.
\end{abstract}


\section{Introduction}
As computer crime first became prevalent, there was discussion of the ``Trojan Horse Defence'' or the ``SODDI Defence'' (Some Other Dude Did It Defence). Some academic literature discussed such cases and how it could be addressed \cite{sepec2012trojan, bowles2015first, lawal2021have}. However, more than two decades since this defence emerged and became prevalent in computer crime, there has been an absence of real-world case studies from forensic investigators describing how this defence can be investigated in a manner capable of acquitting the innocent and convicting the guilty.

This paper finally presents such a case study. The defendant, referred to as \textit{F}, was charged with numerous computer child sexual offences, including making indecent images of children. The defence sought to rely upon this defence.

A forensic investigation conducted by a police detective trained in digital forensics and the author of this paper followed to assist the court in handling the matter and provide the jury with an empirical answer as to the merits of the claims.

This is the first case study of its kind describing how the defence can be addressed in practice, and provides practical lessons and techniques for digital forensic investigators.

\section{Related Work}
There has been limited work in literature on how defences of hacking can be addressed in practice. An early work, \cite{sepec2012trojan} cites a number of historical criminal cases in England \& Wales where the defence has been used successfully and advocates for a multidisciplinary approach. However, as there is no empirical evidence presented for this approach.

Similarly, \cite{bowles2015first} summarises a list of cases from April 2003 to April 2012 where the ``Trojan Horse Defence'' or ``SODDI Defence'' has been used to varying extents. However this is again dated work which comes from a world before the widespread use of mobile phones and does not provide case studies on the forensic science techniques used in such investigations.

\cite{richard2021digital} provides a more contemporaneous source which revisits the ``Trojan Horse Defence'' and argues that given modern malware can be more stealthy (i.e. ``ghostware'') there is a need to incorporate more advanced techniques such as memory forensics. Whilst the paper cites on historical examples, it does not provide empirical evaluation for the approach advocated in real-world criminal cases.

\cite{lawal2021have} provides experimental data from use of a ``USB rubber ducky'', which simulates input/output devices on a computer, to assess which forensic artefacts are left behind. Whilst this paper highlights some forensic artefacts of use, it does not demonstrate such artefacts being used in a real-world criminal case. Similarly, it does not account for forensics being conducted across multiple devices in the possession of a user (e.g. a mobile phone).

There have been advancements in tooling which can identify potential ``Indicators-of-Compromise'' (IOCs) on devices, for example \cite{quetglas2024catalangate} discusses the Mobile Verification Toolkit (``MVT'') developed by Amnesty International, however does not discuss such technology being used in criminal forensics (as the tool was initially designed for counter-espionage purposes).

Therefore, there exists a gap in literature, certainly in computer science, of a forensic case study on how the defence of hacking can be addressed in criminal proceedings.

\section{Background}
Local police received intelligence from the US National Centre for Missing and Exploited Children (NCMEC) via the UK National Crime Agency that a resident was making indecent images of children. In early 2022 a defendant referred to as \textit{F}, was arrested.

\textit{F} was an archived sex offender having previously accepted a caution after a plea deal was reached (but this evidence of bad character could not be later put before the jury to it differing from the previous offending). The defendant’s home was searched, and electronic devices were seized. Amongst those items were an Apple iPhone and a modern Android phone.

Those devices were forensically examined by a detective trained in digital forensics and he confirmed that both devices contained indecent images of children of all categories. In addition, 2 extreme pornographic images and one prohibited image of a child were found on the iPhone.

The defendant claimed their devices were hacked and the author of this paper, Dr Junade Ali, was instructed as an expert witness to comment upon various matters including hacking.

\section{Investigation}
In order to evaluate the credibility of the hacking claims, Dr Ali adopted a staged approach. In the first instance, the possibility of hacking was first considered using paperwork supplied. Secondly, a forensic investigation was conducted on data from the devices to evaluate whether hacking had occurred. Finally, as required by the court, the findings were peer reviewed by the detective and a joint statement was issued.

\subsection{Preliminary Investigation}
The preliminary investigation consisted of a review on paper of the evidence by the parties as to whether hacking was a possibility.

The defence initially sought to rely upon a mobile phone bill and a date when the defendant was putatively in police custody to claim there was unusual network activity on the device. This emerged to be a misunderstanding of how mobile networks roll-up data (i.e. invoicing mobile data usage in aggregate rather than per KB used), and the data itself was consistent with background usage during the time frame the defendant was in custody.

The defence had also claimed, on the basis that the defendant had received compromised credential notifications such as those described in \cite{li2019protocols, ali2017mechanism} that their accounts could have been hacked to plant CSAM (Child Sexual Abuse Material). The report also found this could not have been the case given material was found in an application which require control over a mobile phone number rather than a password. Finally, given there was a conversation soliciting sexual services from a child in another application of which use was admitted, it was not plausible the defendant could not have been aware of the usage of the device for child sexual abuse, even if notifications were disabled.

Finally, the report outlined that given the modern firmwares in use by the devices, zero-day vulnerabilities (with market prices measured in the millions of dollars) would have had to have been used to compromise the devices themselves (to obtain initial access, escape the sandbox and gain privileges). Such hacking was made even more unlikely given two separate modern device operating systems were used (Android and iOS) across the two devices. Indeed, at seizure the firmware was so modern that even with physical access, a number of months had passed until forensic imaging could be done.

\subsection{Forensic Investigation}
The forensic investigation then considered the evidence on the device itself. The detective had already taken forensic images of the devices using Cellebrite UFED and Magnet Forensics Graykey.

Following agreement of a Memorandum of Understanding with the police, forensic images of the devices (alongside a spreadsheet of the paths to any CSAM on the devices) were encrypted and stored on a USB stick.

The encryption key to this USB stick was then emailed to Dr Ali and the memory stick itself was transported from England to Scotland using Royal Mail Special Delivery.

This second report confirmed the presence of CSAM on the device and the material was reviewed to verify that the police had graded the material correctly.

The report then went onto consider the defence of hacking specifically. The techniques used involved manual investigation to verify for the presence of IOCs, and then used the Amnesty International Mobile Verification Toolkit described in \cite{quetglas2024catalangate} to search for any further IOCs.

There were IOCs found on the iPhone, however these exclusively related to those in the \textit{safari\_favicon} module, and Dr Ali concluded that these were the result of the defendant researching cybersecurity and device monitoring tools through web browsing and not connecting to any command-and-control infrastructure. It was clear the iPhone was not jailbroken, which was a prerequisite of such tools; and there were no indicators of backdoors on the device. The investigation also found there was no indication of root access to the Android device, no advanced debugging features enabled or non-standard applications with suspicious permissions.

With the expert opinion excluding evidence, the investigation turned to how such a vast quantity of CSAM (depicting various people) actually appeared on the devices. Absent a contemporaneous memory dump, forensic imaging of the devices and investigation of surrounding chronological forensic artefacts was necessary in order to assess whether there was any indicators of ``ghostware'' performing the actions or it was conducted with user interaction.

Alongside other forensic artefacts demonstrating how the device was being used; the investigation found the bulk of the material was in the cache of the Telegram messenger app and that when Telegram receives a media file it automatically downloads it to the cache (the first few seconds to an autoplay cache, and if the file is below a certain size, the entire file to an app-specific media folder).

The defendant requested to be added to a Telegram group, was added to that group, and as CSAM was shared, the third-party material was also downloaded. This explained why such a vast volume of material was found of the defendant's device even where a specific download button may not have been pressed for each asset.

\subsection{Joint Report}
Dr Ali produced a consolidated forensic report (consolidating both reports into a single report), and as required by the court met with the detective. They produced a joint statement agreeing to the detective's initial ``Streamlined Forensic Report'' outlining the quantity of the material of each type on the devices and Dr Ali's consolidated forensic report. The agreement was wholesale and there were no reservations.

Counsel for the prosecution and the defence then listed the conclusions of Dr Ali's in the agreed facts document, as that both experts had adopted the conclusions.

\section{The Trial}
The trial started in late 2025 in a Crown Court in England and the jury were sworn in. The defendant maintained a not-guilty plea.

The following day the prosecution presented their case. During the opening speech the prosecution barrister relied upon Dr Ali's report to assert the defendant had intentionally downloaded CSAM and was not hacked.

The prosecution called the detective who explained to the jury his forensic investigation of the devices and the material found upon it. When the issue of hacking was raised, the detective gave evidence to the effect that: ``I am not an expert, Dr Ali is the expert and I agree with him.'' The detective further confirmed that there was no evidence that CSAM was planted on the device as a result of hacking. 

The following day the defence was presented. However, the jury had asked two questions related to the operation of the Telegram cache and whether cached would appear in the camera roll. Dr Ali provided answers for both answers to counsel for the parties and this was read by the judge to the jury as an agreed fact.

For the defence, Dr Ali was not required to give oral evidence (although he remained in court throughout the hearing), only the defendant gave evidence on his own behalf. During his evidence he accepted he was not hacked on the basis of the finding of the experts and instead claimed that when he was intoxicated with drugs his friends had used his phones. However during cross-examination, the prosecuting barrister put to the defendant that he had solicited indecent images during a Christmas period when he would have been home and put to the defendant that as each other defence fell away, he had changed his story.

The jury retired later that day, and asked various questions. The final question related to the final paragraph of Dr Ali's final report (which was agreed with by the detective) stating: ``In essence, I believe the likely cause of events here is that the defendant joined a Telegram group sharing CSAM and intentionally copied some CSAM to their devices. The vast majority of the material, however, was downloaded automatically without intention by the Telegram app.''

In English law as criminal matters require both intent (\textit{mens rea}) alongside the act (\textit{actus reus}), the question related to the extent to which they had to be sure the defendant had intended to download the material. Following discussion with counsel, the following day the judge directed the jury that they needed to be sure the defendant had intended to download one of each category of the indecent materials to convict on each count.

In less than an hour after receiving the answer, the defendant was convicted on every count of the indictment.

\section{Discussion}
This case highlights the importance of using a staged hypothesis-testing approach for digital forensics investigations. In this case, this involved firstly identifying reasonable technical lines-of-inquiry to follow on a preliminary basis, a forensic examination which investigated not only artefacts related to the specific defence being advanced but also investigating what the likely actual sequence of events was, and finally seeking peer review on the work undertaken.

Many digital forensic investigations consist of extraction, parsing and reporting of artefacts, however this can often be insufficient.

Indeed, in subsequent work, the author of this paper was instructed in a case where the limitation of this approach in proving a defence was insufficient. In \textit{R v M (2026)}, a defendant had pleaded guilty to numerous child sexual offences, but asserted that they were not the creator of various WhatsApp group chats and therefore believed someone else had created the chats on their devices. A forensic parsing tool used by the police had attributed them to being the creator of the group chats and turning on the disappearing messages functionality. The prosecution had also asserted the defendant was an administrator of these groups on this basis.

Upon instruction a similar investigative approach to one detailed in this paper was undertaken which concluded that neither position was correct. Parsing the artefacts in a different tool (AXIOM by Magnet Forensics), review of underlying SQLite databases and timeline analysis, found that the defendant had not created any of the group chats and was not an administrator of any of them. The likely cause was identified to be a parsing defect in the forensic tool the police were using. This position was ultimately accepted by both sides and the defendant was sentenced on this basis. The police also launched an internal review of the toolchain used in these cases.

Cases such as these highlight a key gap in handling Trojan Horse and SODDI defence cases. Whilst existing approaches typically focus on identifying forensic artefacts of relevance (for example; IOCs and attribution artefacts), automated tooling would not be sufficient to identify reasonable technical lines-of-inquiry and test these hypotheses. This is particularly important where a position in litigation has been advanced on either partial data or non-expert understanding.

Further developments in technology may therefore wish to consider how it can be possible to aid digital forensics experts in identifying such lines-of-inquiry, even where they point away from the position advanced by either or both parties to the litigation.

Validated tooling and techniques in assisting experts in identifying potential hypotheses and testing them holistically can therefore be of benefit in future cases.

\section{Ethics}
This paper describes the methodology used without fine-grained technical specifics or publishing operational detail (e.g. software versions, artefacts, dates, etc.), in order to mitigate the risk of counter-forensics and prevent identification.  Whilst this case was heard in open-court, details are anonymised to prevent jigsaw identification. The author feels it important to publish this work to ensure the risk of miscarriages of justice in future work is mitigated.

\section{Conclusion}
This case study presents a real-world modern case study of addressing the defence of hacking in a sexual offences case in England. Conviction was secured using a layered approach which was first sensitive to the theoretical computer science matrix to identify how a device/accounts could be compromised, followed by a forensic investigation for indicators of compromise on the forensic images of the devices and finally concluded by developing a forensic understanding of what the likely cause of events in the particular case was. In this case, this approach was successful in convincing a jury to the criminal standard, beyond reasonable doubt, that the defendant was not hacked and had committed the offences in question.

Future work may wish to consider how developments in ``pattern-of-life'' identification could be leveraged in a digital forensics context when such a defence is raised. Furthermore, the work highlights how limited the reporting of such cases are. Even amongst legal literature, cases are seldom reported. Expert witnesses are often not told of the outcomes of cases so case studies like the one presented here often go undocumented - limiting knowledge of practical techniques.

Nevertheless, the author hopes the findings here can be of use in shedding light on how the defence of hacking can be evaluated in practice.
\bibliographystyle{IEEEtran} 
\bibliography{bibliography}

\end{document}